\documentclass[twocolumn,amsmath,nofootinbib]{revtex4} 

\usepackage{url}
\usepackage{amssymb}
\usepackage{amsfonts}
\usepackage{amsbsy}
\usepackage{graphicx}
\usepackage{hyperref}
\usepackage{array} 
\usepackage{multirow}
\usepackage{epsfig}
\usepackage{graphics}
\usepackage{amssymb}
\usepackage{amsmath}
\usepackage{comment}
\usepackage{enumerate}
\usepackage{subfigure}
\newcommand{\eqn}[1]{\begin{equation}\begin{split} #1 \end{split}\end{equation}}
\newcommand{\lp}{\left (}
\newcommand{\rp}{\right )}

\newcommand{\ra}{\rightarrow}

\newcommand{\ev}[1]{\left \langle #1 \right \rangle}

\newcommand{\inv}{^{-1}}

\def\ie{{\frenchspacing\it i.e.}}
\def\eg{{\frenchspacing\it e.g.}}

\def\expec#1{\langle#1\rangle}

\def\B{\textbf{B}}
\def\D{\textbf{D}}
\def\M{\textbf{M}}
\def\m{\boldsymbol{\mu}}

\def\spose#1{\hbox to 0pt{#1\hss}}
\def\simlt{\mathrel{\spose{\lower 3pt\hbox{$\mathchar"218$}}
   \raise 2.0pt\hbox{$\mathchar"13C$}}}
\def\simgt{\mathrel{\spose{\lower 3pt\hbox{$\mathchar"218$}}
     \raise 2.0pt\hbox{$\mathchar"13E$}}}
 \def\simpropto{\mathrel{\spose{\lower 3pt\hbox{$\mathchar"218$}}
     \raise 2.0pt\hbox{$\propto$}}}

\def\beq#1{\begin{equation}\label{#1}}
\def\eeq{\end{equation}}
\def\beqa#1{\begin{eqnarray}\label{#1}}
\def\eeqa{\end{eqnarray}}
\def\eq#1{equation~(\ref{#1})}

\def\fig#1{Figure~\ref{#1}}
\def\Fig#1{Figure~\ref{#1}}

\def\Sec#1{Section~\ref{#1}}

\parskip=\medskipamount

\begin{document}
\title{
Criticality in Formal Languages and Statistical Physics\footnote{Published in {\it Entropy}, {\bf 19}, 299 (2017):\\ \url{http://www.mdpi.com/1099-4300/19/7/299}}
%
}
\author{Henry W. Lin and Max Tegmark}
\address{Dept.~of Physics, Harvard University, Cambridge, MA 02138}
\address{Dept.~of Physics \& MIT Kavli Institute, Massachusetts Institute of Technology, Cambridge, MA 02139}
\begin{abstract}
We show that the mutual information between two symbols, as a function of the number of symbols between the two, decays exponentially in any probabilistic regular grammar, but can decay like a power law for a context-free grammar. This result about formal languages is closely related to a well-known result in classical statistical mechanics that there are no phase transitions in dimensions fewer than two. 
It is also related to the emergence of power-law correlations in turbulence and cosmological inflation through recursive generative processes.  
We elucidate these physics connections and comment on potential applications of our results to machine learning tasks like training artificial recurrent neural networks. Along the way, we introduce a useful quantity which we dub the {\it rational mutual information} and discuss generalizations of our claims involving more complicated Bayesian networks. 

\end{abstract}
\date{June 23, 2017}
\vspace{10mm}	

\maketitle

\section{Introduction}
\label{IntroSec}
Critical behavior, where long-range correlations decay as a power law with distance, has many important physics applications ranging from phase transitions in condensed matter experiments to turbulence and inflationary fluctuations in our early Universe. It has important applications beyond the traditional purview of physics as well \cite{bakl,baka,link2001,levi12,tegmark14} including applications to music \cite{manaris, levi12}, genomics \cite{peng, mant94} and human languages \cite{ebeling, ebeling2, altmann, monte02}. 

\begin{figure*}[t]
	\vskip-12mm
	\hglue-20mm\includegraphics[scale=1.0]{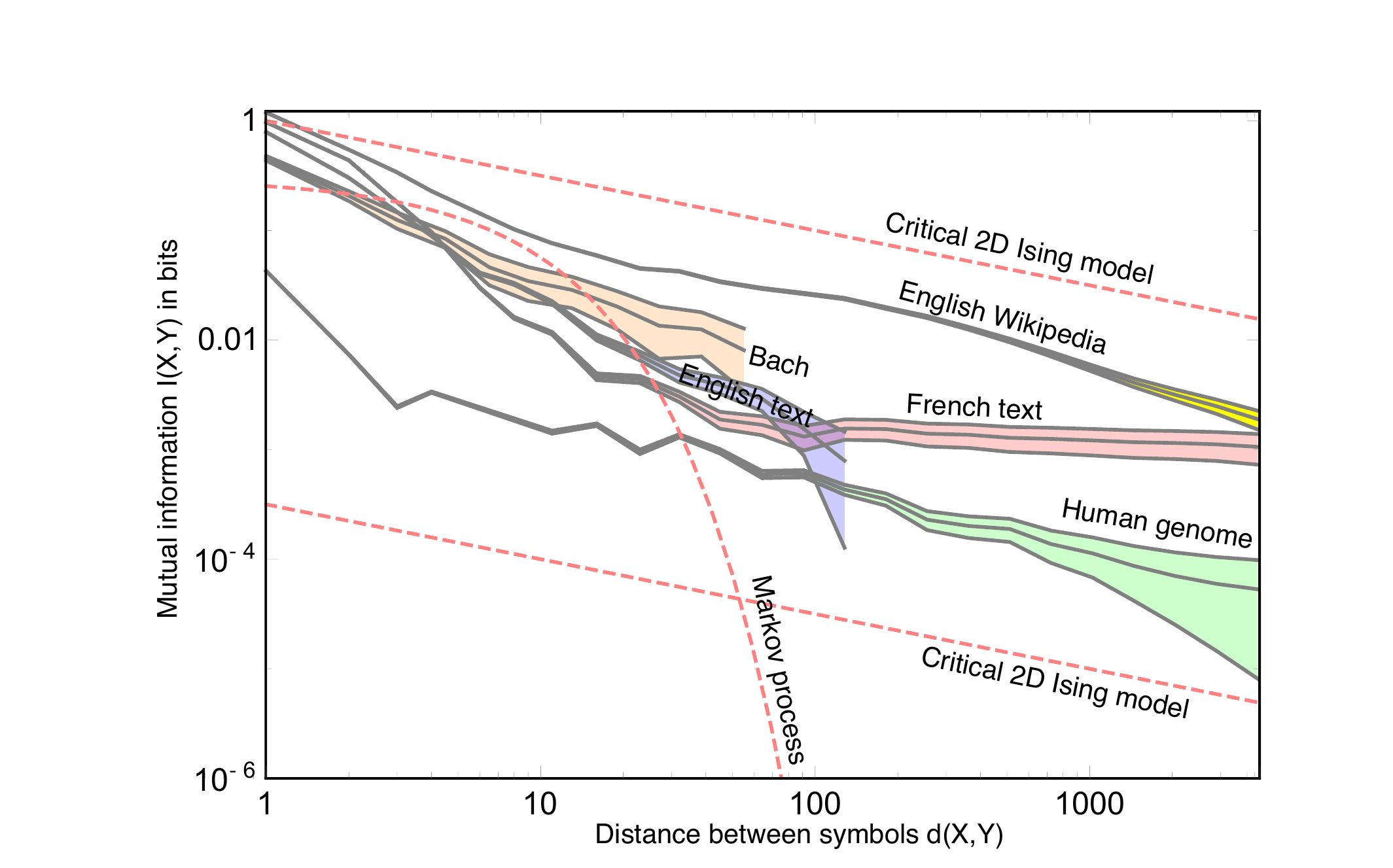}
	\vskip-4mm
	\label{SausageFig}
	\caption{Decay of mutual information with separation. Here the mutual information in bits per symbol is shown as a function of separation $d(X,Y) = |i-j|$, where the symbols $X$ and $Y$ are located at positions $i$ and $j$ in the sequence in question, and shaded bands correspond to $1-\sigma$ error bars. The statistics were computed using a sliding window using an estimator for the mutual information detailed in Appendix D.
	 All measured curves are seen to decay roughly as power laws, explaining why they cannot be accurately modeled as Markov processes --- for which the mutual information instead plummets exponentially (the example shown has $I\propto e^{-d/6}$). The measured curves are seen to be qualitatively similar to that of a famous critical system in physics: a 1D slice through a critical 2D Ising model, where the slope is $-1/2$. The human genome data consists of 177,696,512 base pairs \{A, C, T,G\} from chromosome 5 from the National Center for Biotechnology Information \cite{genome}, with unknown base pairs omitted. The Bach data consists of 5727 notes from Partita No. 2 \cite{bach}, with all notes mapped into a 12-symbol alphabet 
consisting of the 12 half-tones \{C, C\#, D, D\#, E, F, F\#, G, G\#, A, A\#, B\}, with all timing, volume and octave information discarded. The three text corpuses are 100 MB from Wikipedia \cite{hutter} (206 symbols),
the first 114 MB of a French corpus \cite{french} (185 symbols) and  27 MB of English articles from \url{slate.com} (143 symbols). The large long range information appears to be dominated by poems in the French sample and by html-like syntax in the Wikipedia sample.
} 
\end{figure*}

In \fig{SausageFig}, we plot a statistic that can be applied to all of the above examples: the mutual information between two symbols as a function of the number of symbols in between the two symbols \cite{ebeling}. As discussed in previous works \cite{ebeling, deco2012info, altmann}, the plot shows that the number of bits of information provided by a symbol about another drops roughly as a 
power-law\footnote{The power law discussed here should not be confused with another famous power law that occurs in natural languages: Zipf's law \cite{zipf}. Zipf's law implies power law behavior in one-point statistics (in the histogram of word frequencies), whereas we are interested in two-point statistics. In the former case, the power law is in the frequency of words; in the latter case, the power law is in the separation between characters.  
	One can easily cook up sequences which obey Zipf's law but are not critical and do not exhibit a power law in the mutual information. However, there are models of certain physical systems where Zipf's law follows from criticality \cite{lin2016,p2001}.}
 with distance in sequences (defined as the number of symbols between the two symbols of interest) as diverse as the human genome, music by Bach, and text in English and French. Why is this, when so many other correlations in nature instead drop exponentially \cite{kardar}? 

Better understanding the statistical properties of natural languages is interesting not only for geneticists, musicologists and linguists, but also for the machine learning community. Any tasks that involve natural language processing (\eg, data compression,  speech-to-text conversion, auto-correction) exploit statistical properties of language, and can all be further improved if we can better understand these properties, even in the context of a toy model of these data sequences. Indeed, the difficulty of automatic natural language processing has been known at least as far back as Turing, whose eponymous test \cite{turing} relies on this fact. A tempting explanation is that natural language is something uniquely human. But this is far from a satisfactory one, especially given the recent successes of machines at performing tasks as complex and as ``human'' as playing {\it Jeopardy!} \cite{watson}, chess \cite{deepblue}, Atari games \cite{atari} and Go \cite{alphago}. We will show that computer descriptions of language suffer from a much simpler problem that has involves no talk about meaning or being non-human: they tend to get the basic statistical properties wrong.

To illustrate this point, consider Markov models of natural language. From a linguistics point of view, it has been known for decades that such models are fundamentally unsuitable for modeling human language \cite{chomsky59}. However, linguistic arguments typically do not produce an observable that can be used to quantitatively falsify any Markovian model of language. Instead, these arguments rely on highly specific knowledge about the data --- in this case, an understanding of the language's grammar. This knowledge is non-trivial for a human speaker to acquire, much less an artificial neural network. In contrast, the mutual information is comparatively trivial to observe, requiring no specific knowledge about the data, and it immediately indicates that natural languages would be poorly approximated by a Markov/hidden Markov model as we will demonstrate.

Furthermore, the mutual information decay may offer a partial explanation of the impressive progress that has been made by using deep neural networks for natural language processing (see, e.g., \cite{kim15, graves, graves13, collobert, wavenet}). (For recent reviews of deep neural networks, see \cite{sreview, naturereview}.) We will see that a key reason that currently popular recurrent neural networks with long-short-term memory (LSTM) \cite{lstm} do much better is that they can replicate critical behavior, but that even they can be further improved, since they can under-predict long-range mutual information. 

While motivated by questions about natural languages and other data sequences, we will explore the information-theoretic properties of {\it formal} languages. For simplicity, we focus on probabilistic regular grammars and probabilistic context-free grammars (PCFGs). Of course, real-world data sources like English is likely more complex than a context free grammar \cite{shieber}, just as a real-world magnet is more complex than the Ising model. However, these formal languages serve as toy models that capture some aspects of the real data source, and the theoretical techniques we develop for studying these toy models might be adapted to more complex formal languages. Of course, independent of their connection to natural languages, formal languages are also theoretically interesting in their own right and have connections to, e.g., group theory \cite{ani}.

This paper is organized as follows. In \Sec{MarkovSec}, we show how Markov processes exhibit exponential decay in mutual information with scale; we give a rigorous proof of this and other results in a series of appendices. To enable such proofs, we introduce a convenient quantity that we term {\it rational mutual information}, which bounds the mutual information and converges to it in the near-independence limit. 
In \Sec{FractalSec}, we define a subclass of generative grammars and show that they exhibit critical behavior with power law decays. We then generalize our discussion using Bayesian nets and relate our findings to theorems in statistical physics. 
In \Sec{DiscussionSec}, we discuss our results and explain how LSTM RNNs can reproduce critical behavior by emulating our generative grammar model.

\section{Markov implies exponential decay}
\label{MarkovSec}
For two discrete random variables $X$ and $Y$, the following definitions of mutual information are all equivalent:
\eqn{I(X,Y) &\equiv S(X) + S(Y) - S(X,Y) \\ &= D\lp p(XY)\big|\big| p(X) p(Y)\rp \\
	&= \ev{\log_B \frac{P(a,b)}{P(a)P(b)}}\\
	&= \sum_{ab} P(a,b) \log_B \frac{P(a,b)}{P(a)P(b)},
	}
where $S\equiv\expec{-\log_B P}$ is the Shannon entropy \cite{shannon} 
and $D(p(XY)||p(X)p(Y))$ is the Kullback-Leibler divergence \cite{kldiv} between the joint probability distribution and the product of the individual marginals. If the base of the logarithm is taken to be 
$B=2$, then $I(X,Y)$ is measured in bits.
The mutual information can be interpreted as how much one variable knows about the other:
$I(X,Y)$ is the reduction in the number of bits needed to specify for $X$ once $Y$ is specified.
Equivalently, it is the number of encoding bits saved by using the true joint probability $P(X,Y)$ instead of approximating $X$ and $Y$ are independent. It is thus a measure of statistical dependencies between $X$ and $Y$. Although it is more conventional to measure quantities such as the correlation coefficient $\rho$ in statistics and statistical physics, the mutual information is more suitable for generic data, since it does not require that the variables $X$ and $Y$ are numbers or have any algebraic structure, whereas $\rho$ requires that we are able to multiply $X \cdot Y$ and average. Whereas it makes sense to multiply numbers, is meaningless to multiply or average two  characters such as ``!'' and ``?''. 

The rest of this paper is largely a study of the mutual information between two random variables that are realizations of a discrete stochastic process, with some separation $\tau$ in time. More concretely, we can think of sequences $\{X_1, X_2, X_3, \cdots \}$ of random variables, where each one might take values from some finite alphabet. For example, if we model English as a discrete stochastic process and take $\tau = 2$, $X$ could represent the first character (``F'') in this sentence, whereas $Y$ could represent the third character (``r'') in this sentence.

In particular, we start by studying the mutual information function of a Markov process, which is analytically tractable. Let us briefly recapitulate some basic facts about Markov processes (see, \eg, \cite{coverthomas} for a pedagogical review). A Markov process is defined by a matrix $\bf{M}$ of conditional probabilities $M_{ab} = P(X_{t+1} = a|X_{t} = b)$. Such Markov matrices (also known as stochastic matrices) thus have the properties $M_{ab}\ge 0$ and $\sum_{a} M_{ab} = 1$. They fully specify the dynamics of the model:
\eqn{\label{MarkovEq}\textbf{p}_{t+1} = \textbf{M}\,\textbf{p}_t,}
where $\textbf{p}_{t}$ is a vector with components $P(X_t = a)$ that specifies the probability distribution at time $t$. Let $\lambda_i$ denote the eigenvalues of $\M$, sorted by decreasing magnitude:
$|\lambda_1|\ge |\lambda_2|\ge  |\lambda_3|...$ 
All Markov matrices have $|\lambda_i|\le 1$, which is why blowup is avoided when \eq{MarkovEq} is iterated, 
and $\lambda_1=1$, with the corresponding eigenvector giving a stationary probability distribution $\m$ satisfying
$\M\m=\m$.

In addition, two mild conditions are usually imposed on Markov matrices: $\bf{M}$ is {\it irreducible}, meaning that every state is accessible from every other state (otherwise, we could decompose the Markov process into separate Markov processes). Second, to avoid processes like $1 \to 2 \to 1 \to 2 \cdots$ that will never converge, we take the Markov process to be {\it aperiodic}.
It is easy to show using the Perron-Frobenius theorem that being irreducible and aperiodic implies $|\lambda_2|<1$, and therefore that $\m$ is unique.

This section is devoted to the intuition behind the following theorem, whose full proof is given in Appendix A and B. The theorem states roughly that for a Markov process, the mutual information between two points in time $t_1$ and $t_2$ decays exponentially for large separation $|t_2-t_1|$:

{\bf Theorem 1}: Let $\bf{M}$ be a Markov matrix that generates a 
Markov process. 
If $\bf{M}$ is irreducible and aperiodic, then the asymptotic behavior of the mutual information $I(t_1,t_2)$ is exponential decay toward zero for $|t_2 -t_1| \gg 1$ with decay timescale $ \log \frac{1}{|\lambda_2|},$ where $\lambda_2$ is the second largest eigenvalue of $\bf{M}$.
If $\bf{M}$ is reducible or periodic, $I$ can instead decay to a constant; no Markov process whatsoever can produce power-law decay.
Suppose $\bf{M}$ is irreducible and aperiodic so that $\textbf{p}_t \ra \boldsymbol{\mu}$ as $t \ra \infty$ as mentioned above.
This convergence of one-point statistics, \eg, $\textbf{p}_t$, has been well-studied \cite{coverthomas}. However, one can also study higher order statistics such as the joint probability distribution for two points in time. For succinctness, let us write 
$P(a,b)\equiv P(X = a, Y=b) $, where $X = X_{t_1}$ and $Y=X_{t_2}$ and $\tau\equiv |t_2 - t_1| $. 
We are interested in the asymptotic situation where the Markov process has converged to its steady state, 
so the marginal distribution  $P(a)\equiv \sum_b P(a,b)=\mu_a$, independently of time.

If the joint probability distribution approximately factorizes as $P(a,b) \approx \mu_a \mu_b$ for sufficiently large and well-separated times $t_1$ and $t_2$ (as we will soon prove), the mutual information will be small. We can therefore Taylor expand the logarithm from equation (1) around the point $P(a,b) = P(a)P(b)$, giving
\eqn{ I(X,Y) &= \ev{\log_B \lp \frac{P(a,b)}{P(a) P(b)}\rp}\\
&= \ev{\log_B \left[ 1 +  \frac{P(a,b)}{P(a) P(b)} - 1\right]}\\
 &\approx \ev{ \frac{P(a,b)}{P(a) P(b)}-1} \frac{1}{\ln B}
 ={I_R(X,Y)\over\ln B},}
where we have defined the {\it rational mutual information} 
\eqn{I_R \equiv \ev{ \frac{P(a,b)}{P(a) P(b)}-1}.} For comparing the rational mutual information with the usual mutual information, it will be convenient to take $e$ as the base $B$ of the logarithm. We derive useful properties of the rational mutual information in Appendix A. To mention just one, we note that the rational mutual information is not just asymptotically equal to the mutual information in the limit of near-independence, but it also provides a strict upper bound on it: $0 \le I \le I_R$. 

Let us without loss of generality take $t_2>t_1$. Then iterating \eq{MarkovEq} $\tau$ times gives 
$P(b|a)=(\M^\tau)_{ba}$. Since $P(a,b)=P(a)P(b|a)$, we obtain
\eqn{
\label{ir} 
I_R + 1&= \ev{\frac{P(a,b)}{P(a)P(b)}}= \sum_{ab} P(a,b)\frac{P(a,b)}{P(a)P(b)} \nonumber\\
&=  \sum_{ab} \frac{P(b|a)^2 P(a)^2}{P(a) P(b)}
=\sum_{ab} \frac{\mu_a}{\mu_b} [(\M^\tau)_{ba}]^2.
}

We will continue the proof by considering the typical case where the eigenvalues of $\textbf{M}$ are all distinct (non-degenerate) and the Markov matrix is irreducible and aperiodic; we will generalize to the other cases (which form a set of measure zero) in Appendix B. 
Since the eigenvalues are distinct, we can diagonalize $\bf M$ by writing 
\beq{MdiagonalizationEq}
 \M = \B\D\B\inv
\eeq
for some invertible matrix $\B$ and some a diagonal matrix $\D$ whose diagonal elements are the eigenvalues:
$D_{ii}=\lambda_i$. Raising \eq{MdiagonalizationEq} to the power $\tau$ gives 
$\textbf{M}^\tau = \textbf{BD}^\tau\textbf{B}\inv$, \ie, 
\beq{MtauEq}
(\M^\tau)_{ba}= \sum_c  \lambda^\tau_c\>\B_{bc}  (\B\inv)_{ca}.
\eeq
Since $\M$ is non-degenerate, irreducible and aperiodic, $1= \lambda_1 > |\lambda_2| > \cdots > |\lambda_n|$, so all terms except the first in the sum of \eq{MtauEq} decay exponentially with $\tau$, at a decay rate that grows with $c$.
Defining $r = \lambda_3/\lambda_2$, we have
\beqa{MexpansionEq}
(\M^\tau)_{ba} &=& B_{b1} B_{1a}\inv + \lambda_2^\tau\left[ B_{b2} B_{2a}\inv + \mathcal{O}(r^\tau)\right] \nonumber\\
&=& \mu_b +\lambda_2^\tau A_{ba},
\eeqa
where we have made use of the fact that an irreducible and aperiodic Markov process must converge to its stationary distribution for large $\tau$, and we have defined $\textbf{A}$ as the expression in square brackets above, 
satisfying $\lim_{\tau \to \infty} A_{ba} = B_{b2} B_{2a}\inv$. Note that $\sum_{b} A_{ba} = 0$ in order for $\textbf{M}$ to be properly normalized.

Substituting \eq{MexpansionEq} into \eq{ir}  and using the facts that $\sum_{a} \mu_a = 1$ and $\sum_{b} A_{ba} = 0$, we obtain
\eqn{
\label{ir} 
I_R&= \sum_{ab} \frac{\mu_a}{\mu_b} [(\M^\tau)_{ba}]^2-1\\
&=\sum_{ab} \frac{\mu_a}{\mu_b} \lp \mu_b^2 +2 \mu_b \lambda_2^{\tau} A_{ba} +  \lambda_2^{2\tau} A^2_{ba}\rp-1\\
&=\sum_{ab}\lambda_2^{2\tau} \lp \mu_b^{-1}  A_{ba}^2\mu_a  \rp = \mathcal{C} \lambda_2^{2\tau},
}
where the term in the last parentheses is of the form $\mathcal{C} = \mathcal{C}_0 + \mathcal{O}(r^\tau)$.


In summary, we have shown that an irreducible and aperiodic Markov process with non-degenerate eigenvalues cannot produce critical behavior, because the mutual information decays exponentially. In fact, {\it no} Markov processes can, as we show in Appendix B.

To hammer the final nail into the coffin of Markov processes as models of critical behavior, we need to close a final loophole. Their fundamental problem is lack of long-term memory, which can be superficially overcome 
by redefining the state space to include symbols from the past. For example, if the current state is one of $n$ 
and we wish the process to depend on the the last $\tau$ symbols, we can define an expanded state space consisting
of the $n^\tau$ possible sequences of length $\tau$, and a corresponding $n^\tau\times n^\tau$ Markov matrix
(or an $n^\tau\times n$ table of conditional probabilities for the next symbol given the last $\tau$ symbols).
Although such a model could fit the curves in \fig{SausageFig} in theory, it cannot in practice, because 
$\M$ requires way more parameters than there are atoms in our observable universe ($\sim 10^{78}$):
even for as few as $n=4$ symbols and $\tau=1000$, the Markov process involves over 
$4^{1000}\sim 10^{602}$ parameters.
Scale-invariance aside, we can also see how Markov processes fail simply by considering the structure of text.
To model English well, $\M$ would need to correctly close parentheses even if they were opened more than 
$\tau=100$ characters ago, requiring an $\M$-matrix with than $n^{100}$ parameters, where $n>26$ is the number of characters used.

We can significantly generalize Theorem 1 into a theorem about hidden Markov models (HMM). In an HMM, the observed sequence $X_1, \cdots, X_n$ is only part of the picture: there are hidden variables $Y_1, \cdots, Y_n$ that themselves form a Markov chain. We can think of an HMM as follows: imagine a machine with an internal state space $Y$ that updates itself according to some Markovian dynamics. The internal dynamics are never observed, but at each time-step, it also produces some output $Y_i \to X_i$ that form the sequence which we can observe. These models are quite general and are used to model a wealth of empirical data (see, e.g., \cite{rabiner89}).

{\bf Theorem 2}: Let $\bf{M}$ be a Markov matrix that generates the transitions between hidden states $Y_i$ in an HMM. 
If $\bf{M}$ is irreducible and aperiodic, then the asymptotic behavior of the mutual information $I(t_1,t_2)$ is exponential decay toward zero for $|t_2 -t_1| \gg 1$ with decay timescale $ \log \frac{1}{|\lambda_2|},$ where $\lambda_2$ is the second largest eigenvalue of $\bf{M}$.
This theorem is a strict generalization of Theorem 1, since given any Markov process $\mathcal{M}$ with corresponding matrix $\bf{M}$, we can construct an HMM that reproduces the exact statistics of $\mathcal{M}$ by using $\mathcal{M}$ as the transition matrix between the $Y$'s and generating $X_i$ from $Y_i$ by simply setting $x_i = y_i$ with probability 1.

The proof is very similar in spirit to the proof of Theorem 1, so we will just present a sketch here, leaving a full proof to Appendix B. Let $\bf G$ be the Markov matrix that governs $Y_i \to X_i$. To compute the joint probability between two random variables $X_{t_1}$ and $X_{t_2}$, we simply compute the joint probability distribution between $Y_{t_1}$ and $Y_{t_2}$, which again involves a factor of $\textbf{M}^\tau$ and then use two factors of $\bf G$ to convert the joint probability on $Y_{t_1},Y_{t_2}$ to a joint probability on $X_{t_1}, X_{t_2}$. These additional two factors of $\bf G$ will not change the fact that there is an exponential decay given by ${\bf M}^\tau$.

 A simple, intuitive bound from information theory (namely the data processing inequality \cite{coverthomas}) gives $I(Y_{t_1},Y_{t_2}) \ge I(Y_{t_1},X_{t_2}) \ge I(X_{t_1},X_{t_2})$. However, Theorem 1 implies that $I(Y_{t_1},Y_{t_2})$ decays exponentially. Hence $I(X_{t_1},X_{t_2})$ must also decay at least as fast as exponentially.

There is a well-known correspondence between so-called {\it probabilistic regular grammars} \cite{carrasco1994learning} (sometimes referred to as stochastic regular grammars) and HMMs. Given a probabilistic regular grammar, one can generate an HMM that reproduces all statistics and vice versa. Hence, we can also state Theorem 2 as follows:

{\bf Corollary}: No probabilistic regular grammar exhibits criticality.

In the next section, we will show that this statement is not true for context-free grammars.

\section{Power laws from generative grammar} 
\label{FractalSec}


If computationally feasible Markov processes cannot produce critical behavior, then how do such sequences arise? In this section, we construct a toy model where sequences exhibit criticality. In the parlance of theoretical linguistics, our language is generated by a {\it stochastic} or {\it probabilistic context-free grammar} (PCFG) \cite{ginsburg66,booth69,huang71, lari}. We will discuss the relationship between our model and a generic PCFG in Section C.

\subsection{A simple recursive grammar model}
 
We can formalize the above considerations by giving production rules for a toy language $L$ over an alphabet $A$. The language is defined by how a native speaker of $L$ produces sentences: first, she draws one of the $|A|$ characters from some probability distribution $\mu$ on $A$. She then takes this character $x_0$ and replaces it with $q$ new symbols, drawn from a probability distribution $P(b|a)$, where $a \in A$ is the first symbol and $b \in A$ is any of the second symbols. This is repeated over and over. After $u$ steps, she has a sentence of length $q^u$.\footnote{This exponential blow-up is reminiscent of de Sitter space in cosmic inflation. There is actually a much deeper mathematical analogy involving conformal symmetry and $p$-adic numbers that has been discussed \cite{harlow}.}

One can ask for the character statistics of the sentence at production step $u$ given the statistics of the sentence at production step $u-1$. The character distribution is simply 
\eqn{P_u(b) = \sum_a P(b|a) P_{u-1} (a).}
Of course this equation does {\it not} imply that the process is a Markov process when the sentences are read left to right. To characterize the statistics as read from left to right, we really want to compute the statistical dependencies {\it within} a given sequence, \eg, at fixed $u$. 

\begin{figure}[t]
	\vskip-3.4cm
	\includegraphics[scale=0.41]{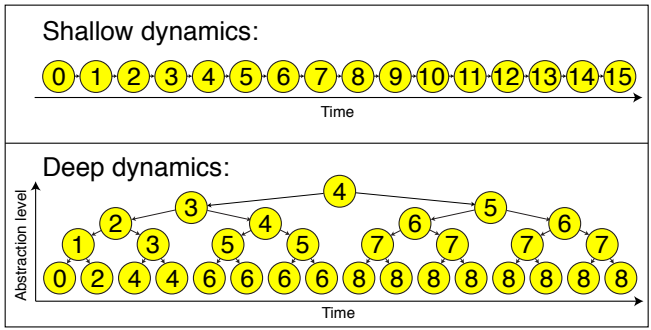}
	\vskip-3.5cm
	\caption{\label{TreeFig}
Both a traditional Markov process (top) and our recursive generative grammar process (bottom) can be represented as Bayesian networks, where
the random variable at each node depends only on the node pointing to it with an arrow.
The numbers show the geodesic distance $\Delta$ to the leftmost node, defined as the smallest number of edges that must be traversed to get there.
Roughly speaking, our results show that for large $\Delta$, the mutual information decays exponentially with $\Delta$ (see Theorem 1 and 2). Since this geodesic distance $\Delta$ grows only logarithmically with the separation in time in a hierarchical generative grammar (the hierarchy creates very efficient shortcuts), the exponential kills the logarithm and we are left with power-law decays of mutual information in such languages.}
\end{figure}

To see that the mutual information decays like a power law rather than exponentially with separation, consider two random variables $X$ and $Y$ separated by $\tau$. One can ask how many generations took place between $X$ and the nearest ancestor of $X$ and $Y$. Typically, this will be about $\log_q \tau$ generations. Hence in the tree graph shown in \fig{TreeFig}, which illustrates the special case $q=2$, the number of edges $\Delta$ between $X$ and $Y$ is about $2 \log_q \tau$. Hence by the previous result for Markov processes, we expect an exponential decay of the mutual information in the variable $\Delta \sim 2 \log_q \tau$. This means that $I(X,Y)$ should be of the form
\eqn{I(X,Y) \sim q^{-\gamma \Delta} =  q^{-2\gamma \log_q \tau} = \tau^{-2\gamma},}
where $\gamma$ is controlled by the second-largest eigenvalue of $\bf G$, the matrix of conditional probabilities $P(b|a)$. But this exponential decay in $\Delta$ is exactly a power-law decay in $\tau$! This intuitive argument is transformed into a rigorous proof in Appendix C.

\subsection{Further Generalization: strongly correlated characters in words}
In the model we have been describing so far, all nodes emanating from the same parent can be freely permuted since they are conditionally independent. In this sense, characters within a newly generated word are uncorrelated. We call models with this property {\it weakly correlated}. There are still arbitrarily large correlations between words, but not inside of words. If a weakly correlated grammar allows $a \to ab$, it must allow for $a \to ba$ with the same probability. We now wish to relax this property to allow for the {\it strongly-correlated} case where variables may not be conditionally independent given the parents. This allows us to take a big step towards modeling realistic languages: in English, {\it god} significantly differs in meaning and usage from {\it dog}.

In the previous computation, the crucial ingredient was the joint probability $P(a,b) = P(X = a, Y = b)$. Let us start with a seemingly trivial remark. This joint probability can be re-interpreted as a conditional joint probability. Instead of $X$ and $Y$ being random variables at {\it specified} sites $t_1$ and $t_2$, we can view them as random variables at randomly chosen locations, conditioned on their locations being $t_1$ and $t_2$. Somewhat pedantically, we write $P(a,b) = P(a,b|t_1,t_2)$. This clarifies the important fact that the only way that $P(a,b|t_1,t_2)$ depends on $t_1$ and $t_2$ is via a dependence on $\Delta(t_1,t_2)$. Hence \eqn{P(a,b|t_1,t_2) = P(a,b|\Delta).}
	This equation is specific to weakly correlated models and does not hold for generic strongly correlated models.

In computing the mutual information as a function of separation, the relevant quantity is the right hand side of equation (7). The reason is that in practical scenarios, we estimate probabilities by sampling a sequence at fixed separation $t_1 -t_2$, corresponding to $\Delta \approx 2 \log_q |t_2 -t_1| + \mathcal{O}(1)$, but varying $t_1$ and $t_2$. (The $\mathcal{O}(1)$ term is discussed in Appendix E). 

Now whereas $P(a,b|t_1,t_2)$ will change when strong correlations are introduced, $P(a,b|\Delta)$ will retain a very similar form.
This can be seen as follows: knowledge of the geodesic distance corresponds to knowledge of how high up the closest parent node is in the hierarchy (see Figure 2). Imagine flowing down from the parent node to the leaves. We start with the stationary distribution $\mu_i$ at the parent node. At the first layer below the parent node (corresponding to a causal distance $\Delta-2$), we get $Q_{rr'} \equiv P(rr') = \sum_i P_S(rr'|i) P(i)$, where the symmetrized probability $P_S = \frac{1}{2} \sum_i [ P(rr'|i)  + P(r'r|i)]$ comes into play because knowledge of the fact that $r,r'$ are separated by $\Delta -2$ gives no information about their order. To continue this process to the second stage and beyond, we only need the matrix $G_{sr} = P(s|r) = \sum_{s'} P_S(s s'|r).$ The reason is that since we only wish to compute the two-point function at the bottom of the tree, the only place where a three-point function is ever needed is at the very top of the tree, where we need to take a single parent into two children nodes. After that, the computation only involves evolving a child node into a grand-child node, and so forth. Hence the overall two-point probability matrix $P(ab|\Delta)$ is given by the simple equation
\eqn{\mathbf{P}(\Delta) = \lp \mathbf{G}^{\Delta/2 - 1}\rp \mathbf{Q} \lp \mathbf{G}^{\Delta/2 - 1}\rp^t.}

As we can see from the above formula, changing to the strongly correlated case essentially reduces to the weakly correlated case where 
\eqn{\mathbf{P}(\Delta) = \lp \mathbf{G}^{\Delta/2}\rp \text{diag}(\boldsymbol{\mu}) \lp \mathbf{G}^{\Delta/2}\rp^t,} except for a perturbation near the top of the tree. We can think of the generalization as equivalent to the old model except for a different initial condition. We thus expect on intuitive grounds that the model will still exhibit power law decay. This intuition is correct, as we will prove in Appendix C. Our result can be summarized by the following theorem:

{\bf Theorem 3} There exist probabilistic context free grammars (PCFGs) such that the mutual information $I(A,B)$ between two symbols $A$ and $B$ in the terminal strings of the language decay like $d^{-k}$, where $d$ is the number of symbols in between $A$ and $B$.

In Appendix C, we give an explicit formula for $k$ as well as the normalization of the power law for a particular class of grammars.



\subsection{Further Generalization: Bayesian networks and context-free grammars}
Just how generic is the scaling behavior of our model? What if the length of the words is not constant? What about more complex dependencies between layers? If we retrace the derivation in the above arguments, it becomes clear that the only key feature of all of our models considered so far is that the rational mutual information decays exponentially with the causal distance $\Delta$:

\eqn{\label{DeltaDecayEq}I_R \sim e^{-\gamma \Delta}.}

This is true for (hidden) Markov processes and the hierarchical grammar models that we have considered above. So far we have defined $\Delta$ in terms of quantities specific to these models; for a Markov process, $\Delta$ is simply the time separation. Can we define $\Delta$ more generically? In order to do so, let us make a brief aside about {\it Bayesian networks}. Formally, a Bayesian net is a directed acyclic graph (DAG), where the vertices are random variables and conditional dependencies are represented by the arrows. Now instead of thinking of $X$ and $Y$ as living at certain times $(t_1, t_2)$, we can think of them as living at vertices $(i,j)$ of the graph.

We define $\Delta(i,j)$ as follows. Since the Bayesian net is a DAG, it is equipped with a partial order $\le$ on vertices. We write $k \le l$ iff there is a path from $k$ to $l$, in which case we say that $k$ is an {\it ancestor} of $l$. We define the $L(k,l)$ to be the number of edges on the shortest directed path from $k$ to $l$. Finally, we define the causal distance $\Delta(i,j)$ to be 
\eqn{\Delta(i,j) \equiv \min_{x \le i, x\le j} L(x,i) + L(x,j).}
It is easy to see that this reduces to our previous definition of $\Delta$ for Markov processes and recursive generative trees (see \fig{TreeFig}).

Is it true that our exponential decay result from \eq{DeltaDecayEq} holds even for a generic Bayesian net? The answer is yes, under a suitable approximation. The approximation is to ignore long paths in the network when computing the mutual information. 
In other words, 
the mutual information tends to be dominated by the shortest paths via a common ancestor, whose length is $\Delta$. This is a generally a reasonable approximation, because these longer paths will give exponentially weaker correlations, so unless the number of paths increases exponentially (or faster) with length, the overall scaling will not change. 

With this approximation, we can state a key finding of our theoretical work. Deep models are important because without the extra ``dimension'' of depth/abstraction, there is no way to construct ``shortcuts'' between random variables that are separated by large amounts of time with short-range interactions; 1D models will be doomed to exponential decay. 
{\it Hence the ubiquity of power laws may partially explain the success of applications of deep learning to natural language processing}. In fact, this can be seen as the Bayesian net version of the important result in statistical physics that there are no phase transitions in 1D \cite{vanhove, cuesta}.

One might object that while the requirement of short-ranged interactions is highly motivated in physical systems, it is unclear why this restriction is necessary in the context of natural languages. Our response is that allowing for allowing for a generic interaction between say $k$-nearest neighbors will increase the number of parameters in the model exponentially with $k$. 

There are close analogies between our deep recursive grammar and more conventional physical systems. 
For example, according to the emerging standard model of cosmology, 
there was an early period of cosmological inflation when density fluctuations get getting added on a 
fixed scale as space itself underwent repeated doublings, combining to produce an excellent approximation to a
power-law correlation function. This inflationary process is simply a special case of 
our deep recursive model (generalized from 1 to 3 dimensions). In this case, the hidden ``depth'' dimension in our model corresponds to cosmic time, and the time parameter which labels the place in the sequence of interest corresponds to space. A similar physical analogy is turbulence in a fluid, where energy in the form of vortices cascades from large scales to ever smaller scales through a recursive process where larger vortices create smaller ones, leading to a scale-invariant power spectrum. 
In both the inflation case and the turbulence case, there is a hierarchical generative process akin to our formal language model (except in three dimensions and with continuous variables), whereby parts of the system generate smaller parts in an essentially Markovian fashion.

There is also a close analogy to quantum mechanics: in equation (13) expresses the exponential decay of the mutual information with geodesic distance through the Bayesian network; in quantum mechanics, the correlation function of a many body system decays exponentially with the geodesic distance defined by the tensor network which represents the wavefunction \cite{evenbly11}.

It is also worth examining our model using techniques from linguistics. A generic PCFG $\mathcal{G}$ consists of three ingredients:
\begin{enumerate}
	\item An alphabet $\mathcal{A} = A \cup T$ which consists of non-terminal symbols $A$ and terminal symbols $T$.
	\item  A set of production rules of the form $a \to B$, where the left hand side $a \in A$ is always a single non-terminal character and $B$ is a string consisting of symbols in $\mathcal{A}$.
	\item Probabilities associated with each production rule $P(a \to B)$, such that for each $a \in A$, $\sum_{B} P(a \to B) = 1$.
\end{enumerate}
It is a remarkable fact that any stochastic-context free grammars can be put in {\it Chomsky normal form} \cite{chomsky59, huang71}. This means that given $\mathcal{G}$, there exists some other grammar $\bar{\mathcal{G}}$ such that all the production rules are either of the form $a \to b c$ or $a \to \alpha$, where $a,b,c \in A$ and $\alpha \in T$ and the corresponding languages $L(\mathcal{G}) = L(\bar{\mathcal{G}})$. In other words, given some complicated grammar $\mathcal{G}$, we can always find a grammar $\bar{\mathcal{G}}$ such that the corresponding statistics of the languages are identical and all the production rules replace a symbol by at most two symbols (at the cost of increasing the number of production rules in $\bar{\mathcal{G}}$).

This formalism allows us to strengthen our claims. Our model with a branching factor $q=2$ is precisely the class of all context-free grammars that are generated by the production rules of the form $a \to b c$. While this might naively seem like a very small subset of all possible context-free grammars, the fact that {\it any} context-free grammar can be converted into Chomsky normal form shows that our theory deals with a generic context-free grammar, except for the additional step of producing terminal symbols from non-terminal symbols. Starting from a single symbol, the deep dynamics of the PCFG in normal form are given by a strongly-correlated branching process with $q=2$ which proceeds for a characteristic number of productions before terminal symbols are produced. Before most symbols have been converted to terminal symbols, our theory applies, and power-law correlations will exist amongst the non-terminal symbols. To the extent that the terminal symbols that are then produced from non-terminal symbols reflect the correlations of the non-terminal symbols, we expect context-free grammars to be able to produce power law correlations.

From our corollary to Theorem 2, we know that regular grammars cannot exhibit power-law decays in mutual information. Hence context-free grammars are the simplest grammars which support criticality, e.g., they are the lowest in the Chomsky hierarchy that supports criticality. Note that our corollary to Theorem 2 also implies that not all context-free grammars exhibit criticality since regular grammars are a strict subset of context-free grammars. Whether one can formulate an even sharper criterion should be the subject of future work.


\section{Discussion}
\label{DiscussionSec}
By introducing a quantity we term rational mutual information, we have proved that hidden Markov processes generically exhibit exponential decay, whereas PCFGs can exhibit power law decays thanks to the ``extra dimension'' in the network. To the extent that natural languages and other empirical data sources are generated by processes more similar to PCFGs than Markov processes, this explains why they can exhibit power law decays.

We will draw on these lessons to give a semi-heuristic explanation for the success of deep recurrent neural networks widely used for natural language processing, and discuss how mutual information can be used as a tool for validating machine learning algorithms.


\subsection{Connection to Recurrent Neural Networks}

While the generative grammar model is appealing from a linguistic perspective, it may appear to have little to do with machine learning algorithms that are implemented in practice. However, as we will now see, this model can in fact be viewed an idealized version of a long-short term memory (LSTM) recurrent neural network (RNN) that is generating (``hallucinating") a sequence. 

\fig{hallucinationFig} shows that an LSTM RNN can reproduce critical behavior. In this example, we trained an RNN (consisting of three hidden LSTM layers of size 256 as described in \cite{graves}) to predict the next character in the 100MB Wikipedia sample known as enwik8 \cite{hutter}.
We then used the LSTM to hallucinate 1 MB of text and measured the mutual information as a function of distance. 
\Fig{hallucinationFig} shows that not only is the resulting mutual information function a rough power law, but it also has a slope that is relatively similar to the original.

We can understand this success by considering a simplified model that is less
powerful and complex than a full LSTM, but retains some of its core features ---
such an approach to studying deep neural nets has proved fruitful in the past (e.g., \cite{saxe}).

\begin{figure}[t]
	\centering
	\vskip-25mm
	\hglue-20mm\includegraphics[width=140mm]{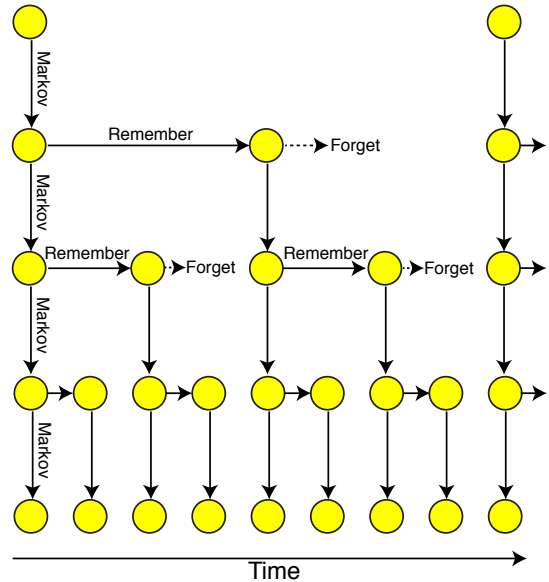}
	\vskip-75mm
	\caption{\label{RNNfig} Our deep generative grammar model can be viewed as an idealization of a long-short term memory (LSTM) recurrent neural net, where the ``forget weights" drop with depth so that the forget timescales grow exponentially with depth. The graph drawn here is clearly isomorphic to the graph drawn in Figure 1. For each cell, we approximate the usual incremental updating rule by either perfectly remembering the previous state (horizontal arrows) or by ignoring the previous state and determining the cell state by a random rule depending on the node above (vertical arrows).}
\end{figure}

\begin{figure}[t]
	\centering
	\subfigure{
	\includegraphics[scale=0.6]{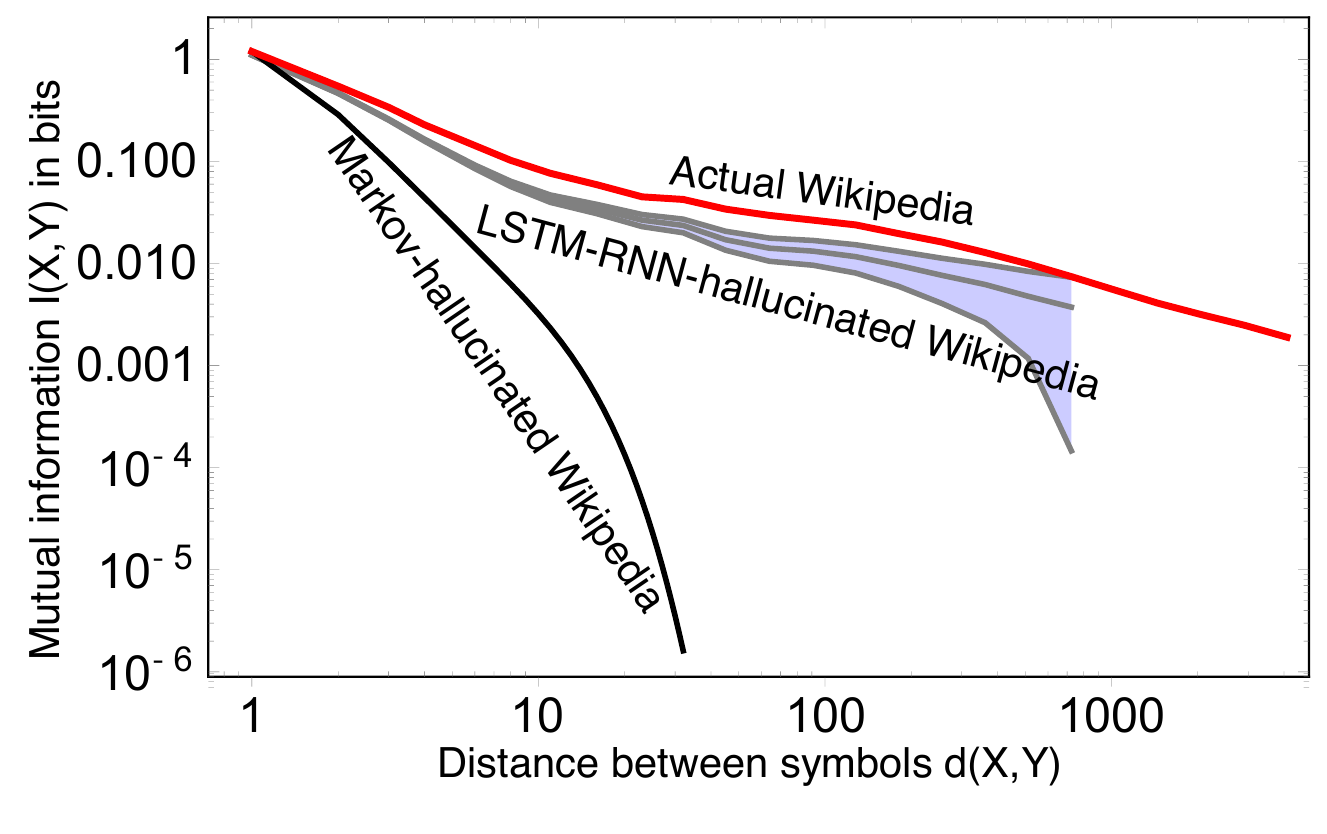}
	}
	\caption{\label{hallucinationFig}Diagnosing different models with by hallucinating text and then measuring the mutual information as a function of separation. The red line is the mutual information of enwik8, a 100 MB sample of English Wikipedia. In shaded blue is the mutual information of hallucinated Wikipedia from a trained LSTM with 3 layers of size 256. We plot in solid black the mutual information of a Markov process on single characters, which we compute exactly. (This would correspond to the mutual information of hallucinations in the limit where the length of the hallucinations goes to infinity). This curve shows a sharp exponential decay after a distance of $\sim 10$, in agreement with our theoretical predictions. We also measured the mutual information for hallucinated text on a Markov process for bigrams, which still underperforms the LSTMs in long-ranged correlations, despite having $\sim 10^3$ more parameters than}
\end{figure}

The usual implementation of LSTMs consists of multiple cells stacked one on top of each other. Each cell of the LSTM (depicted as a yellow circle in Fig. 3) has a state that is characterized by a matrix of numbers $\textbf{C}_t$ and is updated according to the following rule
\eqn{\textbf{C}_t = \textbf{f}_t\circ \textbf{C}_{t-1} + \textbf{i}_t \circ\textbf{D}_t,}
where $\circ$ denotes element wise multiplication, and $\textbf{D}_t = \textbf{D}_t(\textbf{C}_{t-1}, \textbf{x}_t)$ is some function of the input $\textbf{x}_t$ from the cell from the layer above (denoted by downward arrows in \fig{RNNfig}, the details of which do not concern us. Generically, a graph of this picture would look like a rectangular lattice, with each node having an arrow to its right (corresponding to the first term in the above equation), and an arrow from above (corresponding to the second term in the equation). However, if the forget weights $\mathbf{f}$ weights decay rapidly with depth (e.g., as we go from the bottom cell to the towards the top) so that the timescales for forgetting grow exponentially, we will show that a reasonable approximation to the dynamics is given by \fig{RNNfig}.

If we neglect the dependency of $\textbf{D}_t$ on $\textbf{C}_{t-1}$, the forget gate $\textbf{f}_t$ leads to exponential decay of $\textbf{C}_{t-1}$ e.g., $\textbf{C}_{t} = \mathbf{f}^t \circ \textbf{C}_0$; this is how LSTM's forget their past. Note that all operations including exponentiation are performed element-wise in this section only.

 In general, a cell will smoothly forget its past over a timescale of $\sim \log (1/f) \equiv \tau_f$. On timescales $\gtrsim \tau_f$, the cells are weakly correlated; on timescales $\lesssim \tau_f$, the cells are strongly correlated. Hence a discrete approximation to this above equation is the following: 
\eqn{\textbf{C}_t &= \textbf{C}_{t-1}, \text{ for $\tau_f$ timesteps}\\ &= \textbf{D}_t(\textbf{x}_t), \text{ on every $\tau_f+1$ timestep}.}
This simple approximation leads us right back to the hierarchical grammar. The first line of the above equation is labeled ``remember'' in \fig{TreeFig} and the second line is what we refer to as ``Markov,'' since the next state depends only on the previous. Since each cell perfectly remembers its pervious state for $\tau_f$ time-steps, the tree can be reorganized so that it is exactly of the form shown in \fig{RNNfig}, by omitting nodes which simply copy the previous state. Now supposing that $\tau_f$ grows exponentially with depth $\tau_f (\text{layer } i) \propto q \, \tau_f (\text{layer } i+1)$, we see that the successive layers become exponentially sparse, which is exactly what happens in our deep grammar model, identifying the parameter $q$,  governing the growth of the forget timescale, with the branching parameter in the deep grammar model. (Compare \fig{TreeFig} and \fig{RNNfig}.)

\subsection{A new diagnostic for machine learning}


How can one tell whether an neural network can be further improved? 
For example, an LSTM RNN similar to the one we used in \fig{RNNfig} can predict Wikipedia text with a residual entropy $\sim 1.4$ bits/character \cite{graves}, which is very close to the performance of current state of the art custom compression software --- which achieves $\sim 1.3$ bits/character \cite{mahoney}. 
Is that essentially the best compression possible, or can significant improvements be made? 

Our results provide an powerful diagnostic for shedding further light on this question: 
measuring the mutual information as a function of separation between symbols is a computationally efficient way of extracting much more meaningful information about the performance of a model than simply evaluating the loss function, usually given by the conditional entropy $H(X_t|X_{t-1},X_{t-2},\dots)$. 

\Fig{RNNfig} shows that even with just three layers, the LSTM-RNN is able to learn long-range correlations; the slope of the mutual information of hallucinated text is comparable to that of the training set. 
However, the figure also shows that the predictions of our LSTM-RNN are far from optimal.
Interestingly, the hallucinated text shows about the same mutual information for distances $\sim \mathcal{O}(1)$, but significantly less mutual information at large separation. 
Without requiring any knowledge about the true entropy of the input text (which is famously NP-hard to compute), this figure immediately shows that the LSTM-RNN we trained is performing sub-optimally; it is not able to capture all the long-term dependencies found in the training data. 

As a comparison, we also calculated the bigram transition matrix $P(X_3 X_4|X_1 X_2)$ from the data and used it to hallucinate 1 MB of text. Despite the fact that this higher order Markov model needs $\sim 10^3$ more parameters than our LSTM-RNN, it captures less than a fifth of the mutual information captured by the LSTM-RNN even at modest separations $\gtrsim 5$.
This phenomenon is related to a classic result in the theory of formal languages: a context free grammar 

In summary, \fig{RNNfig} shows both the successes and shortcomings of machine learning. On the one hand, LSTM-RNN's can capture long-range correlations much more efficiently than Markovian models; on the other hand, they cannot match the two point functions of training data, never mind higher order statistics!

One might wonder how the lack of mutual information at large scales for the bigram Markov model is manifested in the hallucinated text. Below we give a line from the Markov hallucinations:
\begin{verbatim}
[[computhourgist, Flagesernmenserved whirequotes
or thand dy excommentaligmaktophy as
its:Fran at ||ÃÂ«&lt;If ISBN 088;&ampategor
and on of to [[Prefung]]' and at them rector>
\end{verbatim}

This can be compared with an example from the LSTM RNN:
\begin{verbatim}
Proudknow pop groups at Oxford 
- [http://ccw.com/faqsisdaler/cardiffstwander
--helgar.jpg] and Cape Normans's first 
attacks Cup rigid (AM).
\end{verbatim}

Despite using many fewer parameters, the LSTM manages to produce a realistic looking URL and is able to close brackets correctly \cite{karpathy}, something that the Markov model struggles with.



Although great challenges remain to accurately model natural languages, our results at least allow us to improve on some earlier answers to key questions we sought to address :
\begin{enumerate}
	\item {\it Why is natural language so hard?} The old answer was that language is uniquely human. Our new answer is that at least part of the difficulty is that natural language is a critical system, with long-ranged correlations that are difficult for machines to learn.
	\item {\it Why are machines bad at natural languages, and why are they good?} The old answer is that Markov models are simply not brain/human-like, whereas neural nets are more brain-like and hence better. Our new answer is that Markov models or other 1-dimensional models cannot exhibit critical behavior, whereas neural nets and other deep models (where an extra hidden dimension is formed by the layers of the network) are able to exhibit critical behavior.
	\item {\it How can we know when machines are bad or good?} The old answer is to compute the loss function. Our new answer is to also compute the mutual information as a function of separation, which can immediately show how well the model is doing at capturing correlations on different scales.
\end{enumerate}

Future studies could include generalizing our theorems to more complex formal languages such as Merge Grammars. 

\bigskip
{\bf Acknowledgments:}
This work was supported by the Foundational Questions Institute \url{http://fqxi.org}.
The authors wish to thank Noam Chomsky and Greg Lessard for valuable comments on the linguistic aspects of this work,
Taiga Abe, 
Meia Chita-Tegmark, 
Hanna Field,
Esther Goldberg,
Emily Mu,
John Peurifoi,
Tomaso Poggio,
Luis Seoane,
Leon Shen,
David Theurel, Cindy Zhao, and two anonymous referees for helpful discussions and encouragement, Michelle Xu for help acquiring genome data and the Center for Brains Minds and Machines (CMBB) for hospitality.
\bigskip


\appendix

\section{Properties of rational mutual information}
In this appendix, we prove the following elementary properties of rational mutual information:
\begin{enumerate}
\item {\bf Symmetry}: for any two random variables $X$ and $Y$, $I_R(X,Y) = I_R(Y,X)$. The proof is straightforward:
\eqn{I_R(X,Y)  = \sum_{ab} \frac{P(X=a,Y=b)^2}{P(X=a)P(Y=b)}-1\\ = \sum_{ba} \frac{P(Y=b,X=a)^2}{P(Y=b)P(X=a)}-1 = I_R(Y,X).}

\item {\bf Upper bound to mutual information}: The logarithm function satisfies $\ln(1+x) \le x$ with equality if and only if (iff) $x=0$. Therefore setting $x= \frac{P(a,b)}{P(a)P(b)} - 1$ gives
\eqn{I(X,Y) &= \ev{\log_B \frac{P(a,b)}{P(a)P(b)}} \\
&= {1\over\ln B}\ev{\ln\left[ 1 + \lp \frac{P(a,b)}{P(a) P(b)}-1\rp\right]}\\
&\le {1\over\ln B}\ev{ \frac{P(a,b)}{P(a) P(b)}-1} = {I_R(X,Y)\over\ln B}.\\
}
Hence the rational mutual information $I_R \ge I\ln B$ with equality iff $I = 0$ 
(or simply $I_R \ge I$ if we use the natural logarithm base $B=e$). 

\item {\bf Non-negativity}. It follows from the above inequality that $I_R(X,Y) \ge 0$ with equality iff $P(a,b)=P(a)P(b)$, since $I_R =I=0$ iff $P(a,b) = P(a)P(b)$. Note that this short proof is only possible because of the information inequality $I \ge 0$. From the definition of $I_R$, it is only obvious that $I_R \ge -1$; information theory gives a much tighter bound. Our findings 1-3 can be summarized as follows: 
\eqn{I_R(X,Y) = I_R(Y,X) \ge I(X,Y) \ge 0,}
where both equalities occur iff $p(X,Y) = p(X)p(Y)$. It is impossible for one of the last two relations to be an equality while the other is an inequality.
\item {\bf Generalization}. Note that if we view the mutual information as the divergence between two joint probability distributions, we can generalize the notion of rational mutual information to that of {\it rational divergence}:
\eqn{D_{R}(p||q) = \ev{\frac{p}{q}} - 1,}
	where the expectation value is taken with respect to the ``true'' probability distribution $p$. This is a special case of what is known in the literature as $\alpha$-divergence \cite{amari1985alpha}.
	
	The $\alpha$-divergence is itself a special case of so-called $f$-divergences \cite{morimoto1963markov,cs,al}:
	\eqn{D_f(p|| q) = \sum p_i f(q_i/p_i),}
	where $D_R(p||q)$ corresponds to $f(x) = \frac{1}{x}-1$.
	
	Note that as it is written, $p$ could be any probability measure on either a discrete or continuous space. The above results can be trivially modified to show that $D_R(p||q) \ge D_{KL}(p||q)$ and hence $D_R(p||q) \ge 0$, with equality iff $p = q$.
\end{enumerate}

\section{General proof for Markov processes}
In this appendix, we drop the assumptions of non-degeneracy, irreducibility and non-periodicity made in the main body of the paper where we proved that Markov processes lead to exponential decay.

\subsection{The degenerate case}

First, we consider the case where the Markov matrix $\bf{M}$ has degenerate eigenvalues. In this case, we cannot guarantee that $\bf{M}$ can be diagonalized. However, any complex matrix can be put into Jordan normal form. In Jordan normal form, a matrix is block diagonal, with each $d \times d$ block corresponding to an eigenvalue with degeneracy $d$. These blocks have a particularly simple form, with block $i$ having $\lambda_i$ on the diagonal and ones right above the diagonal. 
For example, if there are only three distinct eigenvalues and $\lambda_2$ is threefold degenerate, the the Jordan form of $\textbf{M}$ would
be
\eqn{\textbf{B}\inv\textbf{MB} = \begin{bmatrix}
1 & 0 & 0 & 0   & 0 \\
 0  & \lambda_2 & 1 & 0   & 0 \\
 0  & 0  & \lambda_2 & 1   & 0 \\ 
 0  & 0  & 0  & \lambda_2 & 0 \\
 0  & 0  & 0  & 0   & \lambda_3
\end{bmatrix}.}
Note that the largest eigenvalue is unique and equal to 1 for all  irreducible and aperiodic $\M$.
In this example, the matrix power $\M^\tau$ is

\eqn{\textbf{B}\inv\textbf{M}^\tau \textbf{B} = \begin{bmatrix}
1 & 0 & 0 & 0   & 0 \\
 0  & \lambda_2^\tau & {\tau \choose 1}\lambda_2^{\tau-1} & {\tau \choose 2}\lambda_2^{\tau-2}   & 0 \\
 0  & 0  & \lambda_2^\tau & {\tau \choose 1}\lambda_2^{\tau-1}    & 0 \\ 
 0  & 0  & 0  & \lambda_2^\tau & 0 \\
 0  & 0  & 0  & 0   & \lambda_3^\tau
\end{bmatrix}.
}

In the general case, raising a matrix to an arbitrary power will yield a matrix which is still block diagonal, with each block being an upper triangular matrix. The important point is that in block $i$, every entry scales $\propto \lambda_i^\tau$, up to a combinatorial factor. Each combinatorial factor grows only polynomially with $\tau$, with the degree of the polynomials in the $i$th block bounded by the multiplicity of $\lambda_i$, minus one.

Using this Jordan decomposition, we can replicate \eq{MexpansionEq} and write

\eqn{M^\tau_{ij} = \mu_i + \lambda_2^\tau A_{ij}.}

There are two cases, depending on whether the second eigenvalue $\lambda_2$ is degenerate or not.
If not, then the equation
\eqn{\lim_{\tau \ra \infty} A_{ij} = B_{i2} B_{2j}\inv}
still holds, since for $i\ge 3$, $(\lambda_i/\lambda_2)^\tau$ decays faster than any polynomial of finite degree. 
On the other hand, if the second eigenvalue is degenerate with multiplicity $m_2$, we instead define $\textbf{A}$ with the combinatorial factor removed:

\eqn{M^\tau_{ij} = \mu_i + {\tau \choose m_2} \lambda_2^\tau  A_{ij}.}
If $m_2 =1$, this definition simply reduces to the previous definition of $\textbf{A}$. With this definition,

\eqn{\lim_{\tau \ra \infty} A_{ij} = \lambda_{2}^{-m_2} B_{i2} B_{(2+m_2) j}\inv,}

Hence in the most general case, the mutual information decays like a polynomial $\mathcal{P}(\tau) e^{-\gamma \tau}$, where $\gamma = 2 \ln \frac{1}{\lambda_2}$. The polynomial is non-constant if and only if the second largest eigenvalue is degenerate. Note that even in this case, the mutual information decays exponentially in the sense that it is possible to bound the mutual information by an exponential.

\subsection{The reducible  case}

Now let us generalize to the case where the Markov process is reducible. A general Markov state space can be partitioned into $m$ subsets, 
\eqn{S = \bigcup_{i=1}^m S_i,} 
where elements in the same partition communicate with each other: it is possible to transition from $i \to j$ and $j \to i$ for $i,j \in S_i$. 

In general, the set of partitions will be a finite directed acyclic graph (DAG), where the arrows of the DAG are inherited from the Markov chain. Since the DAG is finite, after some finite amount of time, almost all the probability will be concentrated in the ``final'' partitions that have no outgoing arrows and almost no probability will be in the ``transient'' partitions. Since the statistics of the chain that we are interested are determined by running the chain for infinite time, they are insensitive to transient behavior, and hence we can ignore all but the final partitions. (The mutual information at fixed separation is still determined by averaging over all (infinite) time steps.)


Consider the case where the initial probability distribution only has support on one of the $S_i$. Since states in $S_j \ne S_i$ will never be accessed, the Markov process (with this initial condition) is identical to an irreducible Markov process on $S_i$. Our previous results imply that the mutual information will exponentially decay to zero.

Let us define the random variable $Z = f(X)$, where $f(x \in S_i) = S_i$. For a general initial condition, the total probability within each set $S_i$ is independent of time. This means that the entropy $H(Z)$ is independent of time. Using the fact that $H(Z|X) = H(Y|X) = 0$, one can show that 
\eqn{I(X,Y) = I(X,Y|Z) + H(Z),}
where $I(X,Y|Z) = H(X|Z) - H(Y|X,Z)$ is the conditional mutual information. Our previous results then imply that the conditional mutual information decays exponentially, whereas the second term $H(Z) \le \log m$ is constant.
In the language of statistical physics, this is an example of topological order which leads to constant terms in the correlation functions; here, the Markov graph of $\M$ is disconnected, so there are $m$ degenerate equilibrium states.

\subsection{The periodic case}

If a Markov process is periodic, one can further decompose each final partition. It is easy to check that the period of each element in a partition must be constant throughout the partition. It follows that each final partition $S_i$ can be decomposed into cyclic classes $S_{i1}, S_{i2}, \cdots, S_{id}$, where $d$ is the period of the elements in the partition in $S_i$. The arguments in the previous section with $f(x\in S_{ik}) = S_{ik}$ then show that the mutual information again has two terms, one of which exponentially decays, the other of which is constant.

\subsection{The $n>1$ case}

The following proof holds only for order $n=1$ Markov processes, but we can easily extend the results for arbitrary $n$. Any $n=2$ Markov process can be converted into an $n=1$ Markov process on pairs of letters $X_1 X_2$. Hence our proof shows that $I(X_1 X_2, Y_1 Y_2)$ decays exponentially. But for any random variables $X,Y$, the data processing inequality \cite{coverthomas} states that $I(X,g(Y)) \le I(X,Y)$, where $g$ is an arbitrary function of $Y$. Letting $g(Y_1 Y_2) = Y_1$, and then permuting and applying $g(X_1, X_2) = X_1$ gives
\eqn{I(X_1 X_2,Y_1 Y_2) \ge I(X_1 X_2, Y_1) \ge I(X_1, Y_1).}
Hence, we see that $I(X_1,Y_1)$ must exponentially decay. The preceding remarks can be easily formalized into a proof for an arbitrary Markov process by induction on $n$.


\subsection{The detailed balance case}

This asymptotic relation can be strengthened for a subclass of Markov processes which obey a condition known as detailed balance. This subclass arises naturally in the study of statistical physics \cite{gardiner}. For our purposes, this simply means that there exist some real numbers $K_m$ and a symmetric matrix $S_{ab} = S_{ba}$ such that
\eqn{ M_{ab} = e^{K_a/2} S_{ab} e^{-K_b/2}.}
Let us note the following facts. (1) The matrix power is simply $\lp M^\tau\rp _{ab}= e^{K_a/2} \lp S^\tau\rp_{ab} e^{-K_b/2}$. 
(2) By the spectral theorem, we can diagonalize $S$ into an orthonormal basis of eigenvectors, which we label as $v$ (or sometimes $w$), e.g., $Sv=\lambda_i v$ and $v \cdot w = \delta_{vw}$. Notice that 
$$\sum_n M_{ab} e^{K_n/2} v_n = \sum_n e^{K_m/2} S_{mn} v_n = \lambda_i e^{K_m/2}v_m.$$

Hence we have found an eigenvector of $M$ for every eigenvector of $S$. Conversely, the set of eigenvectors of $S$ forms a basis, so there cannot be any more eigenvectors of $M$. This implies that all the eigenvalues of $M$ are given by $P^{v}_m = e^{K_m/2} v_m$, and the eigenvalues of $P^v$ are $\lambda_i$. In other words, $M$ and $S$ share the same eigenvalues. 

(3) $\mu_a = \frac{1}{Z}e^{K_a}$ is an eigenvector with eigenvalue 1, and hence is the stationary state:
\eqn{	\sum_b M_{ab} \mu_b = \frac{1}{Z}\sum_b e^{(K_a + K_b)/2} S_{ab} \\
	= \frac{1}{Z} e^{K_a} \sum_b e^{K_b/2} S_{ba} e^{-K_a/2} = \mu_a \sum_b M_{ba} = \mu_a.
}
The previous facts then let us finish the calculation:
\eqn{	\ev{ \frac{P(a,b)}{P(a) P(b)}} = \sum_{ab} \lp e^{K_a} \lp S^\tau \rp_{ab}^2 e^{-K_b}\rp \lp e^{K_b-K_a}\rp \\
	= \sum_{ab} \lp e^{K_a} \lp S^\tau \rp_{ab}^2 e^{-K_b}\rp \lp e^{K_b-K_a}\rp  \\
	= \sum_{ab} \lp S^\tau \rp_{ab}^2 = ||S^\tau||^2.
}
Now using the fact that $||A ||^2  = \text{tr} \lp A^T A\rp$ and is therefore invariant under an orthogonal change of basis, we find that
\eqn{ 
\ev{ \frac{P(a,b)}{P(a) P(b)}} = \sum_i |\lambda_i|^{2\tau}.}
Since the $\lambda_i$'s are both the eigenvalues of $M$ and $S$, and since $M$ is irreducible and aperiodic, there is exactly one eigenvalue $\lambda_1 = 1$, and all other eigenvalues are less than one. Altogether,
\eqn{ I_R(t_1,t_2) = \ev{ \frac{P(a,b)}{P(a) P(b)}} -1 = \sum_{i = 2} |\lambda_i|^{2\tau}. }

Hence one can easily estimate the asymptotic behavior of the mutual information if one has knowledge of the spectrum of $M$. We see that the mutual information exponentially decays, with a decay scale time-scale given by the second largest eigenvalue $\lambda_2$:
\eqn{ \tau_\text{decay}^{-1} = 	{2 \log \frac{1}{\lambda_2}}.}

\subsection{Hidden Markov Model}
In this subsection, we generalize our findings to hidden Markov models and present a proof of Theorem 2. If we have a Bayesian network of the form $W \leftarrow X \rightarrow X \rightarrow Y \rightarrow Z$, one can show that $I(W,Z) \le I(X,Y)$ using arguments similar to the proof of the data processing inequality. Hence if $I(X,Y)$ decays exponentially, $I(W,Z)$ should also decay exponentially. In what follows, we will show this in greater detail.

Based on the considerations in the main body of the text, the joint probability distribution between two visible states $X_{t_1}, X_{t_2}$ is given by
\eqn{P(a,b) = \sum_{cd} G_{bd} \left[ \lp M^\tau \rp_{dc} \mu_c\right] G_{ac},}
where the term in brackets would have been there in an ordinary Markov model and the two new factors of $G$ are the result of the generalization. Note that as before, $\mu$ is the stationary state corresponding to $\bf M$. We will only consider the typical case where $\bf M$ is aperiodic, irreducible, and non-degenerate; once we have this case, the other cases can be easily treated by mimicking our above proof for or ordinary Markov processes. Using equation (7) and defining $\textbf{g} = \textbf{M} \boldsymbol{\mu}$ gives
\eqn{P(a,b) &= \sum_{cd} G_{bd} \left[ \lp M^\tau \rp_{dc} \mu_c\right] G_{ac}
\\
&= g_a g_b + \lambda_2^\tau \sum_{cd} \lp G_{bd}  A_{dc} \mu_c G_{ac}\rp.}
Plugging this in to our definition of rational mutual information gives
\eqn{I_R + 1 &= \sum_{ab} \frac{P(a,b)^2}{g_a g_b}\\
&=\sum_{ab} \lp g_a g_b + \lambda_2^\tau \sum_{cd} G_{bd} A_{dc} \mu_c G_{ac} \rp \\
&+ \lambda_2^{2\tau} \mathcal{C}\\
&= 1 + \lambda_2^\tau \sum_{cd} A_{dc} \mu_c + \lambda_2^{2\tau} \mathcal{C}\\
&= 1+ \lambda_2^{2\tau} \mathcal{C},
}
where we have used the facts that $\sum_{i} G_{ij} = 1$, $\sum_{i}A_{ij} = 0$, and as before $\mathcal{C}$ is asymptotically constant. This shows that $I_R \propto \lambda_2^{2\tau}$ exponentially decays.
\section{Power laws for generative grammars}
In this appendix, we prove that the rational mutual information decays like a power law for a sub-class of generative grammars. We proceed by mimicking the strategy employed in the above appendix. Let $\bf G$ be the linear operator associated with the matrix $P_{b|a}$, the probability that a node takes the value $b$ given that the parent node has value $b$. We will assume that $\bf G$ is irreducible and aperiodic, with no degeneracies. From the above discussion, we see that removing the degeneracy assumption does not qualitatively change things; one simply replaces the procedure of diagonalizing $\bf G$ with putting ${\bf G}$ in Jordan normal form.

Let us start with the weakly correlated case. In this case,
\eqn{\label{joint:general}P(a, b) &= \sum_r \mu_r \lp G^{\Delta/2}\rp_{ar} \lp G^{\Delta/2}\rp_{br},}
since as we have discussed in the main text, the parent node has the stationary distribution $\boldsymbol{\mu}$ and $\textbf{G}^{\Delta/2}$ give the conditional probabilities from transitioning from the parent node to the nodes at the bottom of the tree that we are interested in. We now employ our favorite trick of diagonalizing $\bf{G}$ and then writing
\eqn{(G^{\Delta/2})_{ij} = \mu_i + \lambda_2^{\Delta/2} A_{ij},}
which gives 
\eqn{P(a, b) &= \sum_r \mu_r \lp \mu_a + \lambda_2^{\Delta/2} A_{ar} \rp \lp \mu_b + \lambda_2^{\Delta/2} A_{br} \rp,\\
&=\sum_{r} \mu_r \lp \mu_a \mu_b + \mu_a \epsilon  A_{br} + \mu_b \epsilon A_{ar} + \epsilon^2 A_{ar} A_{br}\rp\\}

where we have defined $\epsilon = \lambda_2^{\Delta/2}$. Now note that $\sum_r A_{ar}\mu_r = 0$, since $\boldsymbol{\mu}$ is an eigenvector with eigenvalue 1 of $\mathbf{G}^{\Delta/2}.$ Hence this simplifies the above to just

\eqn{P(a, b) &= \mu_a \mu_b + \epsilon^2 \sum_{r} \mu_r  A_{ar} A_{br}.\\}
From the definition of rational mutual information,
and employing the fact that $\sum_i A_{ij} =0$ gives
\eqn{I_R +1 & \approx \sum_{ab} \frac{\lp \mu_a \mu_b + \epsilon^2 \sum_{r} \mu_r  A_{ar} A_{br} \rp ^2}{\mu_a \mu_b}\\
& = \sum_{ab}  \left[ \mu_a \mu_b + \epsilon^4 N_{ab}^2 \right], \\
& = 1 + \epsilon^4||\textbf{N}||^2,
}
where $N_{ab} \equiv (\mu_a \mu_b)^{-1/2} \sum_r \mu_r A_{ar} A_{br}$ is a symmetric matrix and $|| \cdot||$ denotes the Frobenius norm. Hence
\eqn{I_R = \lambda_2^{2\Delta} ||S||^2.}


Let us now generalize to the strongly correlated case. As discussed in the text, the joint probability is modified to

\eqn{\label{joint:generalgen}P(a, b) &= \sum_{rs} Q_{rs} \lp G^{\Delta/2-1}\rp_{ar} \lp G^{\Delta/2-1}\rp_{bs},}

where $Q$ is some symmetric matrix which satisfies $\sum_r Q_{rs} = \mu_s$. We now employ our favorite trick of diagonalizing $\bf{G}$ and then writing
\eqn{(G^{\Delta/2})_{ij} = \mu_i + \epsilon A_{ij},}
where $\epsilon \equiv \lambda_2^{\Delta/2 - 1}$. This gives 
\eqn{P(a, b) &= \sum_{rs} Q_{rs} \lp \mu_a + \epsilon A_{ar} \rp \lp \mu_b + \epsilon A_{bs} \rp,\\
&= \mu_a \mu_b + \sum_{rs} Q_{rs} \lp \mu_a \epsilon  A_{bs} + \mu_b \epsilon A_{ar} + \epsilon^2 A_{ar} A_{bs}\rp.\\
&= \mu_a \mu_b + \sum_s \mu_a \epsilon A_{bs} \mu_s + \sum_r \mu_b \epsilon A_{ar} \mu_r \\
&+ \epsilon^2\sum_{rs} Q_{rs} A_{ar}A_{bs}\\
&= \mu_a \mu_b + \epsilon^2\sum_{rs} Q_{rs} A_{ar}A_{bs}.}
Now defining the symmetric matrices $R_{ab} \equiv \sum_{rs} Q_{rs} A_{ar}A_{bs} \equiv \lp \mu_a \mu_b \rp^{1/2} N_{ab}$, and noting that $\sum_a R_{ab} = 0$, we have
\eqn{I_R +1 &= \sum_{ab} \frac{\lp \mu_a \mu_b + \epsilon^2 R_{ab}\rp^2}{\mu_a \mu_b}\\
& = \sum_{ab}  \left[ \mu_a \mu_b + \epsilon^4 N_{ab}^2 \right], \\
& = 1 + \epsilon^4||\textbf{N}||^2,
}
which gives
\eqn{I_R = \lambda_2^{2\Delta-4} ||\textbf{N}||^2.}
In either the strongly or the weakly correlated case, note that $\textbf{N}$ is asymptotically constant. We can write the second largest eigenvalue $|\lambda_2|^2 = q^{-k_2/2}$, where $q$ is the branching factor,
\eqn{ I_R \propto q^{-\Delta k_2/2 } \simpropto q^{-k_2 \log_q|i-j|} = \mathcal{C} |i-j|^{-k_2}. }
Behold the glorious power law! We note that the normalization $\mathcal{C}$ must be a function of the form $\mathcal{C} = m_2 f(\lambda_2, q)$, where $m_2$ is the multiplicity of the eigenvalue $\lambda_2$. We evaluate this normalization in the next section.

As before, this result can be sharpened if we assume that $\bf G$ satisfies detailed balance $G_{mn} = e^{K_m/2} S_{mn} e^{-K_n/2}$ where $\bf S$ is a symmetric matrix and $K_n$ are just numbers. Let us only consider the weakly correlated case. By the spectral theorem, we diagonalize $S$ into an orthonormal basis of eigenvectors $v$. As before, $G$ and $S$ share the same eigenvalues. Proceeding, 
%
%
%
\eqn{\label{joint:detailedbalance} P(a, b) = \frac{1}{Z} \sum_v \lambda_v^{\Delta} v_a v_b e^{\lp K_a + K_b\rp/2},}
where $Z$ is a constant that ensures that $P$ is properly normalized.
%
%
Let us move full steam ahead to compute the rational mutual information:
\eqn{
\sum_{ab} \frac{P(a,b)^2}{P(a)P(b)} \\ 
&=  \sum_{ab} e^{-\lp K_a + K_b\rp} \lp \sum_{v} \lambda_v^{\Delta} v_a v_b e^{\lp K_a + K_b\rp/2} \rp^2\\ 
&= \sum_{ab} \lp \sum_v \lambda_v^{\Delta} v_a v_b \rp^2.}
This is just the Frobenius norm of the symmetric matrix $H \equiv \sum_v \lambda_v^{\Delta} v_a v_b $! The eigenvalues of the matrix can be read off, so we have

\eqn{ I_R(a,b) = \sum_{i = 2} |\lambda_i|^{2\Delta}. }

Hence we have computed the rational mutual information exactly as a function of $\Delta$.In the next section, we use this result to compute the mutual information as a function of separation $|i-j|$, which will lead to a precise evaluation of the normalization constant $\mathcal{C}$ in the equation  
\eqn{ I(a,b) \approx \mathcal{C} |i-j|^{-k_2}. }
%

\subsection{Detailed evaluation of the normalization}

For simplicity, we specialize to the case $q=2$ although our results can surely be extended to $q>2$. Define $\delta = \Delta/2$ and $d = |i-j|$. We wish to compute the expected value of $I_R$ conditioned on knowledge of $d$. By Bayes rule, $p(\delta|d) \propto p(d|\delta)p(\delta)$. Now $p(d|\delta)$ is given by a triangle distribution with mean $2^{\delta-1}$ and compact support $(0, 2^\delta)$. On the other hand, $p(\delta) \propto 2^\delta$ for $\delta \le \delta_\text{max}$ and $p(\delta) = 0$ for $\delta \le 0$ or $\delta > \delta_\text{max}$. This new constant $\delta_\text{max}$ serves two purposes. First, it can be thought of as a way to regulate the probability distribution $p(\delta)$ so that it is normalizable; at the end of the calculation we formally take $\delta_\text{max} \to \infty$ without obstruction. Second, if we are interested in empirically sampling the mutual information, we cannot generate an infinite string, so setting $\delta_\text{max}$ to a finite value accounts for the fact that our generated string may be finite.

We now assume $d \gg 1$ so that we can swap discrete sums with integrals. We can then compute the conditional expectation value of $2^{-k_2 \delta}$. This yields
\eqn{I_R \approx \int_{0}^\infty 2^{-k_2 \delta} P(d|\delta) \, \text{d} \delta = \frac{ \left(1-2^{-k_2}\right)d^{-k_2}}{k_2 (k_2+1) \log (2)}, }
or equivalently, \eqn{\mathcal{C}_{q=2} = \frac{1-|\lambda_2|^4}{k_2 (k_2+1)}\frac{1}{\log 2}.}

It turns out it is also possible to compute the answer without making any approximations with integrals:
\eqn{I_R = \frac{2^{-\left(k_2+1\right) \left\lceil \log _2(d)\right\rceil} \left(\left(2^{k_2+1}-1\right) 2^{\left\lceil \log _2(d)\right\rceil }-2 d \left(2^{k_2}-1\right)\right)}{2^{k_2+1}-1}.
}

\begin{figure}[t]
	\vskip-30mm
	\hglue-20mm\includegraphics[width=120mm]{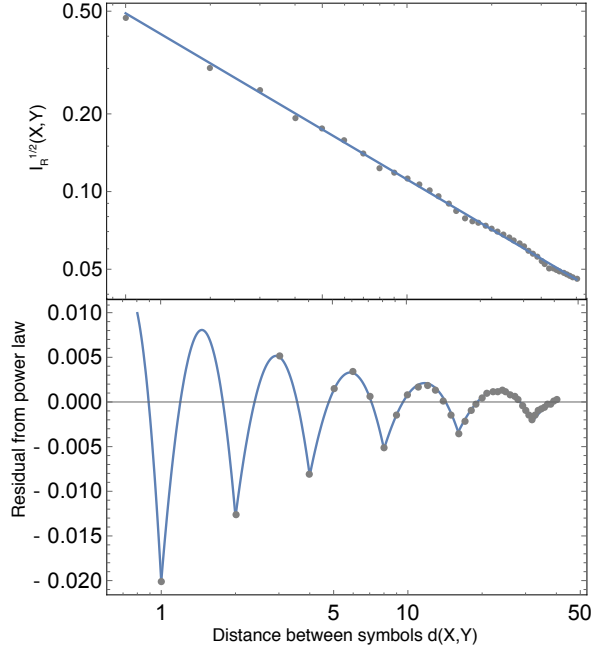}
	\vskip-40mm
	\caption{\label{ApproximationFig}Decay of rational mutual information with separation for a binary sequence from a numerical simulation with probabilities $p(0|0) = p(1|1) = 0.9$ and a branching factor $q=2$. The blue curve is {\it not} a fit to the simulated data but rather an analytic calculation. The smooth power law displayed on the left is what is predicted by our ``continuum'' approximation. The very small discrepancies (right) are not random but are fully accounted for by more involved exact calculations with discrete sums.}
\end{figure}

The resulting predictions are compared in figure \fig{ApproximationFig}.

\section{Estimating (rational) mutual information from empirical data}

Estimating mutual information or rational mutual information from empirical data is fraught with subtleties.

It is well known that a naive estimate of the Shannon entropy obtained $\hat{S} = -\sum_{i=1}^K \frac{N_i}{N} \log \frac{N_i}{N}$ is biased,
generally underestimating the true entropy from finite samples. For example, 
We use the estimator advocated by Grassberger \cite{grassberger}:
\eqn{\hat{S} = \log N - \frac{1}{N}\sum_{i=1}^K N_i \psi(N_i),}
where $\psi(x)$ is the digamma function, $N = \sum N_i$, and $K$ is the number of characters in the alphabet. The mutual information estimator can then be estimated by $\hat{I}(X,Y)= \hat{S}(X) + \hat{S}(Y) - \hat{S}(X,Y)$. The variance of this estimator is then the sum of the variances 
\eqn{\text{var}(\hat{I}) = \text{varEnt}(X) + \text{varEnt}(Y) + \text{varEnt}(X,Y),}
where the varEntropy is defined as
\eqn{\text{varEnt}(X) = \text{var}\lp - \log p(X),\rp}
where we can again replace logarithms with the digamma function $\psi$. The uncertainty after $N$ measurements is then $\approx \sqrt{\text{var}(\hat{I})/N}$.

To compare our theoretical results with experiment in Fig. 4, we must measure the rational mutual information for a binary sequence from (simulated) data. For a binary sequence with covariance coefficient $\rho(X,Y) = P(1,1) - P(1)^2$, the rational mutual information is
\eqn{I_R(X,Y) = \lp \frac{\rho(X,Y)}{P(0)P(1)} \rp^2.}
This was essentially calculated in \cite{liw} by considering the limit where the covariance coefficient is small $\rho \ll 1$. In their paper, there is an erroneous factor of 2. To estimate covariance $\rho(d)$ as a function of $d$ (sometimes confusingly referred to as the correlation function), we use the unbiased estimator for a data sequence $\{x_1, x_2, \cdots x_n\}$:

\eqn{\hat{\rho}(d) = \frac{1}{n-d-1}\sum_{i=1}^{n-d} \lp x_i -\bar{x} \rp \lp x_{i+d} - \bar{x}\rp.}

However, it is important to note that estimating the covariance function $\rho$ by averaging and then squaring will generically yield a biased estimate; we circumvent this by simply estimating $I_R(X,Y)^{1/2} \propto \rho(X,Y)$.

\bibliography{fractal}

\end{document}